\documentstyle[twocolumn,epsf]{jpsj}
\def \virg{\;\;,}

\def \kf{k_{\rm F}}

\def \td{t_{\rm d}}

\def \W0c{W_{0 {\rm c}}}
\def \W0B{W_{0 {\rm B}}}
\def\ggs{\buildrel\textstyle > \over {\hbox{\raise0.2ex\hbox{$\sim$}}}}
\def\lls{\buildrel\textstyle < \over {\hbox{\raise0.2ex\hbox{$\sim$}}}}
\def\gsim{\,\lower0.75ex\hbox{$\ggs$}\,}
\def\lsim{\,\lower0.75ex\hbox{$\lls$}\,}
\def \sgn{{\rm sgn}}
\def\runtitle{
 Quarter-filled  spin  density  wave  states  
 with   long-range Coulomb interaction
  }
\def\runauthor
{
 Y. Tomio, Y. Suzumura 
}

\title{   
 Quarter-filled  spin density wave states 
  with  long-range \\ Coulomb interaction\footnote{%
submitted to J. Phys. Chem. Solids}
  }

\author{
Y. Tomio$^a$, Y. Suzumura$^{a,b,}$\footnote{Corresponding author.  
 Tel.:  81-52-789-2437. 
 fax:   81-52-789-2932. 
E-mail address: e43428a@nucc.cc.nagoya-u.ac.jp}
}

\inst{
$^{a}$Department of Physics, Nagoya University, Nagoya 464-8602, JAPAN \\ 
$^{b}$CREST, Japan Science and Technology Corporation (JST) \\
}

\recdate{\hspace{3.5cm} }

\recdate{\hspace{30mm}}

\abst
{  
 Spin density wave (SDW) states at quarter-filling, 
  which coexist with   charge density wave (CDW) states,  
     have been examined where the critical temperature 
 is calculated  for an extended Hubbard model  
 with long range repulsive interactions.
 Within the mean-field theory, it is shown that 
   the first order transition   
    occurs  with decreasing temperature  
       for interactions   
     located around  the boundary between 
                 SDW state and CDW state. 
}

\kword
{
A. Organic compound, D. Spin-density waves, D. Charge-density waves,
D. Phase transition
  }

\begin{document}
\sloppy
\maketitle
\section{Introduction}
 Bechgaard salts ( (TMTSF)$_2$X and (TMTTF)$_2$X ),
  which are known as low dimensional organic conductors, 
     exhibit spin density wave (SDW) states at low temperatures.
\cite{Jerome,Yamaji}
The SDW states exhibit several  unconventional properties
 associated with charge density wave (CDW). 
  The recent X-ray experiment on  (TMTSF)$_2$PF$_6$ salt
\cite{Pouget,Kagoshima}  
  has shown that 2$\kf$-SDW coexists with  2$\kf$-CDW 
      at temperatures just below  the onset temperature of SDW  
         where   $\kf$ denotes a Fermi momentum.
The  coexistence of SDW and CDW has been studied theoretically 
  in terms of the mean-field theory  
     at the absolute zero temperature.
  By taking into account repulsive interactions of 
    both on-site and  nearest-neighbor sites and dimerization, 
 it has been   demonstrated   that  
 2$\kf$-SDW coexists with  4$\kf$-CDW.\cite{Seo} 
Further 
  the  coexistence of 2$\kf$-SDW and 2$\kf$-CDW
  has been found  by adding    
   the  next-nearest-neighbor repulsive interaction.
\cite{Kobayashi}

 In the present paper, by extending the previous calculations,
\cite{Seo,Kobayashi,Tomio}
 we study  
   if such a long range Coulomb 
 interaction  results in the coexistence    
    even at the onset temperature of the SDW state, as found in the 
experiment.
\cite{Pouget,Kagoshima} 
 
\section{Formulation}
 We examine a one-dimensional  
    extended Hubbard model with interactions of on-site ($U$),  
     nearest-neighbor ($V$), 
       next-nearest-neighbor ($V_2$) sites and 
          dimerization energy ($\td$). The Hamiltonian is  expressed as
\cite{Kobayashi} 
\begin{eqnarray}   \label{Hamiltonian} 
	H  &=&  - \sum_{\sigma=\uparrow,\downarrow}  
         \sum_{j=1}^{N}  
         \left( t - (-1)^j \td  \right) 
	 \left( C_{j\sigma}^\dagger  C_{j+1,\sigma} + h.c. \right) 
                      \nonumber    \\
     & &  
\hspace{-8mm}
           {}  +  U \sum_{j=1}^{N}
               n_{j \uparrow} n_{j \downarrow}  
	     + \sum_{j=1}^{N} 
	     V_{1j}  n_{j} n_{j+1}   
	     + V_2 \sum_{j=1}^{N} n_{j} n_{j+2} ~.
\end{eqnarray}
 The quantity $C^\dagger_{j\sigma}$  denotes 
  the creation  operator of the electron
    at the $j$-th  site  and $V_{1j}=V - (-1)^j \delta V$ where 
 $\delta V$ also comes from dimerization 
 and 
    $n_j=n_{j \uparrow}+n_{j \downarrow}$ with 
          $n_{j\sigma}=C^\dagger_{j\sigma}C_{j\sigma}$.   
 Quantities $t$, $k_{\rm B}$ and lattice constant are taken as unity. 
  Due to the  quarter-filled band with the Fermi wave vector 
 $\kf=\pi/4$,  
 order parameters with $ m=1,2,3$   are calculated self-consistently by 
\begin{eqnarray}
   S_{mQ_0} &=& \frac{1}{N} \sum_{\sigma=\uparrow,\downarrow}  
                \sum_{-\pi < k \leq \pi } \sgn (\sigma)  
                \left<C_{k\sigma}^\dagger C_{k+mQ_0,\sigma}
                 \right>_{\rm MF} ,  
		                  \label{OPS} \nonumber \\ \\   
   D_{mQ_0} &=& \frac{1}{N} \sum_{\sigma=\uparrow,\downarrow}  
                \sum_{-\pi < k \leq \pi }   
                \left<C_{k\sigma}^\dagger C_{k+mQ_0,\sigma}
                \right>_{\rm MF}     
                         \virg        \label{OPD}  
\end{eqnarray}%
where
$Q_0 = 2\kf$,
 $S_0=0$, $D_0=1/2$, $S_{Q_0}=S^*_{3Q_0}$, $D_{Q_0}=D^*_{3Q_0}$, 
$S_{2Q_0}=S^*_{2Q_0} \equiv S_2$ and  $D_{2Q_0}=D^*_{2Q_0} \equiv D_2$.
  In  Eqs. (\ref{OPS}) and (\ref{OPD}), 
    $S_{1}(\equiv |S_{Q_0}|)$,  $S_{2}$,
      $D_{1}(\equiv |D_{Q_0}|)$ and   $D_{2}$
         correspond to the amplitudes for  2$\kf$-SDW, 4$\kf$-SDW,
           2$\kf$-CDW and  4$\kf$-CDW respectively.
From eqs. (\ref{OPS}) and (\ref{OPD}), 
   the free energy per site 
     with the quantity  $U/16+V/4+V_2/4$ subtracted   is  given by
\begin{eqnarray}   \label{FMF}
 F_{\rm MF}  &=&  -\frac{T}{ N} \sum_{\sigma}   
         \sum_{0< k \leq Q_0 } \sum_{n=1}^{4}
         \ln \Bigl( 1 + \exp [ (E_{n\sigma}(k)-\mu)/T ] \Bigr)  
 \nonumber  \\
        & &
\hspace{-9mm}
     {}  + 
       U \left[ - \frac{1}{8} -\frac{1}{2} \left( 
       {|D_{Q_0}|}^2 - {|S_{Q_0}|}^2 \right) 
     - \frac{1}{4} \left( D_{2Q_0}^{2} - S_{2Q_0}^{2} 
       \right) \right]
 \nonumber  \\ 
        & &
\hspace{-9mm}
     {}  + 
       V  \left(  -\frac{1}{2} + D^2_{2Q_0}  \right) 
     + {\rm i} \delta V  \left(  D^2_{Q_0} - D^{*2}_{Q_0}  \right)
 \nonumber  \\
        & &
\hspace{-9mm} 
     {}  + V_2 \left( -\frac{1}{2} + 2 {|D_{Q_0}|}^2 
       - D^2_{2Q_0} \right) + \frac{\mu}{2}  
 \virg
\end{eqnarray}
where $T$ is temperature and 
$\mu$ is a chemical potential determined by $D_0 = 1/2$. 
 In eq. (\ref{FMF}), 
   $E_{n \sigma}$ is the eigen value for the mean-filed Hamiltonian,
\cite{Tomio} 
\begin{eqnarray}   \label{MFH} 
   H_{\rm MF} & = & 
	  \sum_{\sigma=\uparrow,\downarrow}   
         \sum_{-\pi< k \leq \pi }  
     \Bigl[ \left( \varepsilon_k + \frac{U}{4} + V + V_2 \right)
         C_{k\sigma}^\dagger C_{k\sigma}   
 \nonumber  \\
        & &
\hspace{-16mm}
         +  \Bigl(  \Delta_{Q_0 \sigma}
         C_{k+Q_0 ,\sigma}^\dagger  C_{k\sigma} + h.c.  \Bigr)
        + \Delta_{2Q_0 \sigma} 
         C_{k\sigma}^{\dagger} C_{k+2Q_0,\sigma} \Bigr] ,
\end{eqnarray}
 where  $\varepsilon_k=-2t \cos k$, 
 $\Delta_{Q_0\sigma}=\bigl( U/2 -2V_2 \bigr) D_{Q_0}
    + 2{\rm i} \delta V D^*_{Q_0} - \sgn(\sigma) US_{Q_0}/2$  
 and 
$\Delta_{2Q_0\sigma}=\bigl( U/2 -2V + 2V_2 \bigr) D_{2Q_0}
    - \sgn(\sigma) U S_{2Q_0}/2 - 2 {\rm i} \td \sin k$. 
 Here we note an excess free energy,  
 $\delta F$, which is obtained by expanding   
 $F_{\rm MF}$ in terms of order parameters. 
Since there are two kinds of coupling for 
$D_2 S_1^2$ and   $S_1 S_2 D_1$   
 as found in the next section, 
 the relevant expressions for $\delta F$ are written  as   
\begin{eqnarray}
   \label{DELF1}
 \delta F & = &  B_{S_1} S_1^2  +  B_{D_2} D_2^2
        +  G_1 D_2 S_1^2 + \cdots    \virg  \\  
    \label{DELF2}
 \delta F & = & B_{S_1} S_1^2  +  B_{S_2} S_2^2
        +  B_{D_1} D_1^2  +  G_2 S_1 S_2 D_1  + \cdots \virg
 \nonumber \\ 
\end{eqnarray} 
 where $G_1$ and $G_2$ are coupling constants. 
 For small $B_{X}$, one obtains   
     $B_{X} \propto (T-T_{X}^0)$ 
          with  $T_{X}^0$ ($X = S_1, D_2$ and $D_1$) being   
              the onset temperature 
    for $S_1$-state, $D_2$-state and $D_1$-state.

\section{Phase diagram at finite temperatures}

  We examine  SDW  states at finite temperatures, 
 by calculating 
  eqs. (\ref{OPS}), (\ref{OPD}) and  (\ref{FMF}). 
   The numerical calculation is performed 
    by using   $t_b=t-\td$ and  $V_a=V+\delta V$  
  with the fixed  $\td/t$= 1/21, $U/t=3.81$ and   $\delta V/V$ = 1/9.
\cite{Kobayashi}

%
\begin{figure}[t]
\epsfysize=7.6cm
\hspace{6mm}
   \epsffile{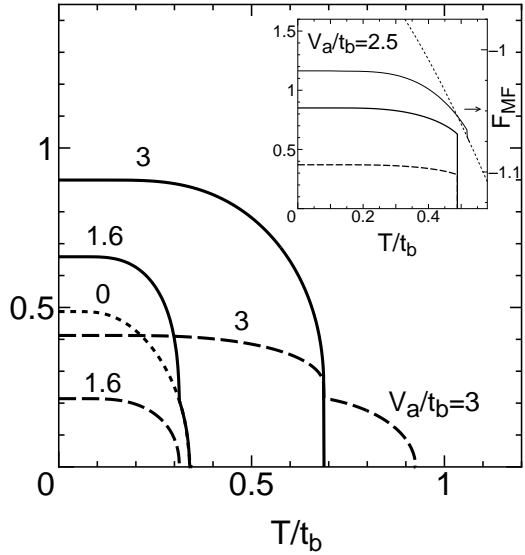}
\vspace{-2mm}
\caption{
 The $T/t_b$-dependence of order parameters,
   $S_1$ (solid and dotted curves) and $D_2$ (dashed curve)  
     for  $V_a/t_b$ =0,  1.6  and 3 with the fixed $V_2/V_a$ = 0. 
 In the inset,  corresponding  $S_1$ and $D_2$  for $V_a/t_b$ = 2.5
    (left axis) is shown with 
  free energies (right axis)  
    for $V_a/t_b = 2.5$ (thin solid curve) 
    and  the normal state (thin dotted curve). 
}
\end{figure}
%
In Fig. 1, $T$-dependence of order parameters 
 for $V_2=0$ and  $V_a/t_b$ =0, 1.6 and 3 is shown where 
   solid curve and dashed curve correspond to $S_1$ and  $D_2$,
   respectively  
  and  $S_2=D_1=0$. 
 The dotted  curve with $V_a$ =0  denotes 
 a  conventional  2$\kf$-SDW state.  
 There are two kinds of phase transitions for $V_a/t_b$ = 1.6 (3)
   where  pure $S_1$ state ($D_2$ state) is obtained at high temperatures 
     with  $0.314 < T/t_b <0.341$  ($0.688 < T/t_b <0.924$)
 while  a coexistent state of $S_1$ and $D_2$ is obtained 
   at low temperature with  $T/t_b < 0.314$ ($T/t_b < 0.688$).
 The inset denotes a first order transition 
  into a coexistent state of $S_1$ and $D_2$, 
   which occurs  at $T/t_b$=0.488 for  $V_a/t_b$ = 2.5.  
  The first order transition temperature is estimated 
 by comparing $F_{\rm MF}$ of   eq.(\ref{FMF}).
%
\begin{figure}[t]
\epsfysize=7.5cm
\hspace{3mm}
   \epsffile{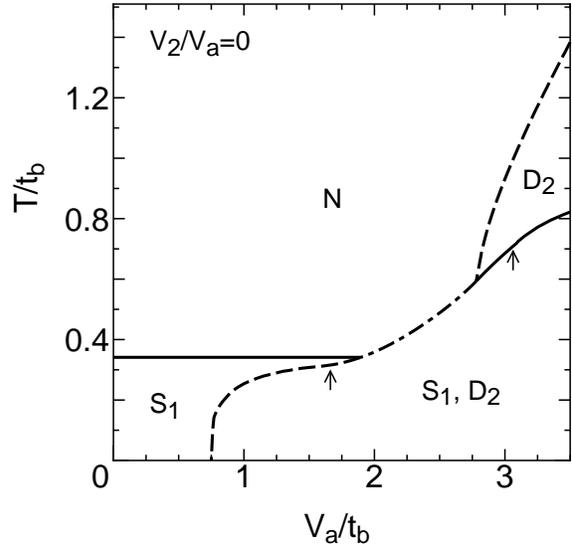}
\vspace{-2mm}
\caption{
 Phase diagram on the plane of $V_a/t_b$ and $T/t_b$ with $V_2/V_a$ = 0
   where $S_1$, $D_2$  and $N$ correspond to 
   2$\kf$-SDW, 4$\kf$-CDW and normal state, respectively.
 In the interval region between  two values ($V_a/t_b$) 
 shown by  two  arrows, 
 a   first order transition takes place  with increasing temperature from 
 the ground state. 
}
\end{figure}
%
Based on these results, phase diagram on the plane of $V_a/t_b$ and 
 $T/t_b$ for $V_2=0$ is shown in Fig. 2, 
 where $S_1 \not= 0$ ($D_2  \not= 0$) at temperatures below 
  the solid curve (dashed curve). 
 With decreasing temperature, 
 the second order transition into $S_1$ state occurs for $V_a/t_b < 0.75$
  while the second order transition into  $S_1$ state is followed 
     by the second (first) order  transition into the coexistent state 
       of $S_1$ and $D_2$ for  $0.75 < V_a/t_b < 1.67$ 
         ($1.67 < V_a/t_b < 1.89$). 
It is noticeable that 
  the first order transition from the normal ($N$) state 
    into the coexistent state of $S_1$ and $D_2$ 
     takes place on the  dash-dotted curve
          with  $1.89 < V_a/t_b < 2.78$.
 Further the second order transition into  $D_2$ state is followed 
 by the first (second) order  transition into the coexistent state 
 of $S_1$ and $D_2$ for  $2.78 < V_a/t_b < 3.07$   ($3.07< V_a/t_b$ ).
Equation (\ref{DELF1}) indicates a fact that  
  the first order transition originates 
    in the third term with a coefficient $G_1$ 
     and that the second order transition from $S_1$ state to 
        the coexistent state of  $S_1$ and $D_2$ is due to 
          $G_1 =0$ in the pure $S_1$ state.  
Thus the first order transition for $V_2=0$ is attributable to
     the third term of eq. (\ref{DELF1}). 

%
\begin{figure}[t]
\epsfysize=7.6cm
\hspace{6mm}
   \epsffile{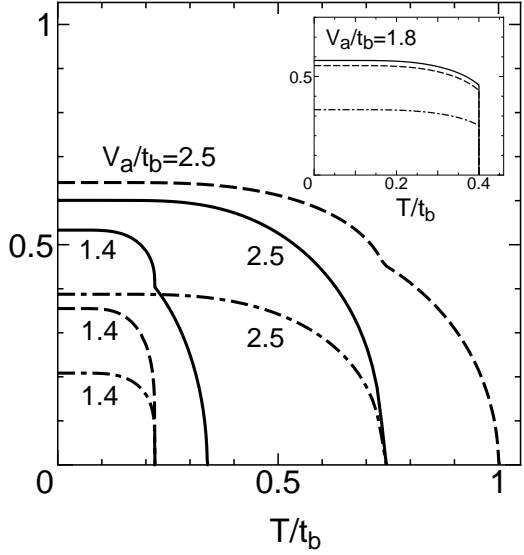}
\vspace{-2mm}
\caption{
 The $T/t_b$-dependence of order parameters,
  $S_1$ (solid curve), $S_2$ (dash-dotted curve) 
    and $D_1$ (dashed curve)  for  $V_a/t_b$ =1.4 and 2.5
        with the fixed $V_2/V_a$ = 1. 
 The inset shows  order parameters for   $V_a/t_b$ = 1.8, where 
   $S_1$, $S_2$,  and $D_1$ vanish  followed by a jump 
     at a critical temperature. 
}
\end{figure}
%
Next we examine  another case of $V_2/V_a= 1$.
In Fig. 3,  $T$-dependence of order parameters is shown   
 with some choices of $V_a/t_b$ where  the solid curve, 
 dashed curve and dash-dotted curve correspond to 
 $S_1$, $D_1$ and $S_2$, respectively.   
 For $V_a/t_b$ = 1.4 (2.5), 
   pure $S_1$ state ($D_1$ state) is obtained at high temperatures 
      with   $0.223 < T/t_b <0.341$  ($0.746 < T/t_b <1.00$)
 while  a coexistent state of $S_1$, $D_1$ and $S_2$ is obtained 
     at low temperature with  $T/t_b < 0.223$ ($T/t_b < 0.746$).
 The inset denotes a first order transition, which occurs  
    at $T/t_b$=0.4 for  $V_a/t_b$ = 1.8.
%
\begin{figure}[t]
\epsfysize=7.5cm
\hspace{3mm}
   \epsffile{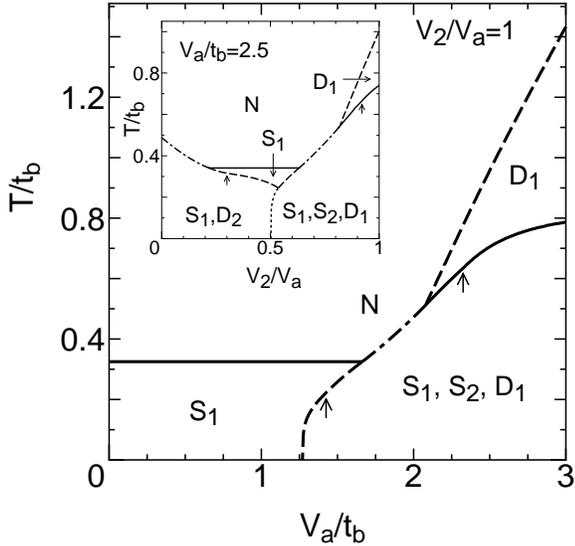}
\vspace{-2mm}
\caption{
 Phase diagram on the plane of $V_a/t_b$ and $T/t_b$ with $V_2/V_a$ = 1.
 The notations are the same as in Fig. 2, where 
  $S_2$ and $D_1$ denote 4$\kf$-SDW and 2$\kf$-CDW, respectively.  
The inset denotes a phase diagram on the plane of $V_2/V_a$ and $T/t_b$
 with $V_a/t_b =$ 2.5.  
}
\end{figure}
%
 For $V_2/V_a=1.0$,  the phase diagram  
 on the plane of $V_a/t_b$ and $T/t_b$  is shown in Fig. 4,  
 where $S_1 \not= 0$ ($D_1  \not= 0$) below 
  the solid curve (dashed curve).  
 For $V_a/t_b < 1.26$, the second order transition into $S_1$ state
   occurs while the second order transition into  $S_1$ state is followed 
     by the second (first) order  transition into the coexistent state 
      of $S_1$, $D_1$ and $S_2$ for  $1.26 < V_a/t_b < 1.43$ 
         ($1.43 < V_a/t_b < 1.66$). 
 A salient feature is  
  the first order transition on the dash-dotted curve, where 
   the normal ($N$) state moves  
    into the coexistent state of $S_1$, $D_1$ and $S_2$ 
     in the interval region of $1.66 < V_a/t_b < 2.08$. 
    The second order transition into  $D_1$ state is followed 
     by the first (second) order  transition into the coexistent state 
     of $S_1$, $D_1$ and $S_2$
    for  $2.08 < V_a/t_b < 2.32$  ($2.32< V_a/t_b$ ).
  The first order transition originates 
     in the third term with a coefficient $G_2$ of eq. (\ref{DELF2}). 
It should be noticed that the existence of  $S_2$ 
  is crucial to  obtain a direct transition 
 from $N$ state to the coexistent of $S_1$ and $D_1$. 
 Actually, for $V_a/t_b=2.5$ and $V_2/V_a=0.7$, 
  the mean-field calculation with $S_2=0$
   leads to a second order transition from $N$ state to pure $S_1$ state,
\cite{Ogata}
  while the corresponding  calculation in the presence of $S_2$
     exhibits a first order transition   
         into a coexistent state.  

%
\begin{figure}[t]
\epsfysize=7.5cm
\hspace{3mm}
   \epsffile{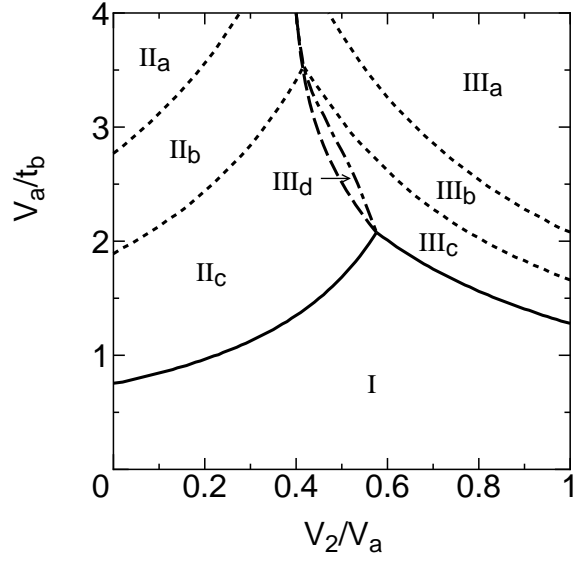}
\vspace{-2mm}
\caption{
 Phase diagram on the plane of $V_2/V_a$ and $V_a/t_b$. 
The solid curve (dashed curve) shows the boundary for $T=0$,
    between the  pure $S_1$ state and the coexistent state 
     (between two kinds of coexistent state).\cite{Tomio}
 The dotted curve corresponds to a boundary at 
     the  critical temperature 
  where it is obtained from   the intersection of 
    the solid curve  and the dashed curve in Figs. 2 and 4.
 The respective regions together with the dash-dotted curve 
   are depicted in the main text. 
}
\end{figure}
%
 Base on these results, a phase diagram on a plane of 
  $V_2/V_a$ and $V_a/t_b$ is shown in Fig. 5. 
  With decreasing temperatures, we obtain 
   the following phase transitions. 
In the region I, a transition from $N$ state into pure $S_1$ state 
  appears   while the region  II$_a$ (II$_c$) shows 
 the successive transition given by 
  $ N $   $\rightarrow$  $D_2$    $\rightarrow$   
     $S_1$ and  $D_2$  
( $ N $   $\rightarrow$  $S_1$    $\rightarrow$   
          $S_1$ and  $D_2$ ).  
 The first order transition from 
 $N$ state into a coexistent state of $S_1$ and $D_2$
   is obtained  in the region II$_b$.
  In the region  III$_a$ (III$_c$), there is 
    the successive transition given by 
   $ N $   $\rightarrow$  $D_1$    $\rightarrow$   
     $S_1$, $D_1$ and $S_2$ 
( $ N $   $\rightarrow$  $S_1$    $\rightarrow$   
    $S_1$, $D_1$ and $S_2$).   
The first order transition from 
 $N$ state into a coexistent state of $S_1$, $D_1$ and $S_2$
    is obtained  in the region III$_b$ while  the region III$_d$ 
    corresponds to  
     the  transition given by 
  $ N   \rightarrow    S_1   \rightarrow   S_1, D_2  
 \rightarrow S_1, D_1$ and  $S_2$.   
 The region  III$_d$ is understood by an  example 
  shown in the inset of Fig. 4  on the plane of $V_2/V_a$ and $T/t_b$,  
where  the corresponding transition is obtained 
  for $ 0.5 < V_2/V_a < 0.54$ for $V_a/t_b = 2.5$.  


By studying a model at quarter-filling 
  with long range Coulomb interactions, 
  we have found  that, 
   with decreasing temperature,   
a first order transition 
 from $N$  state into the coexistent state of SDW and CDW 
  occurs. Since  there is  a reasonable range of parameters, 
 it is considered that 
 the present result  of the first order transition could be  relevant
  to    the coexistent state of $2\kf$-SDW and 2$\kf$-CDW found in 
   the X-ray  experiment.
\cite{Pouget,Kagoshima}

\vspace{1cm}
{\bf Acknowledgment}  \\
The authors thank  M. Ogata  and K. Yonemitsu 
 for useful discussions. 
 This work was partially supported by  a Grant-in-Aid 
 for Scientific  Research  from the Ministry of Education, 
Science, Sports and Culture (Grant No.09640429), Japan.



\end{document}